\documentclass[12pt,letterpaper]{article}
\usepackage[utf8]{inputenc}

\usepackage{graphicx,array}
\usepackage{url}
\usepackage{color}
\usepackage{latexsym}
\usepackage{amsthm}
\usepackage{amsmath}
\usepackage{amssymb}
\usepackage{amsfonts}
\usepackage[numbers,sort&compress]{natbib}
\usepackage{bm}
\usepackage{slashed}
\usepackage{mathrsfs}
\usepackage{enumerate}
\usepackage{tikz}
\usepackage{siunitx}
\usepackage{mdframed}
\usepackage{setspace}  
\usepackage{esvect}

\usepackage{tcolorbox}%

\usepackage{hyperref} 
\hypersetup{
    colorlinks=true,       
    linkcolor=red,          
    citecolor=blue,        
    filecolor=magenta,      
    urlcolor=blue           
}
\usepackage[all]{hypcap} 
\usepackage{multirow}
\usepackage{multicol}

\usepackage{natbib}
\newcommand{\nc}{\newcommand}  
\newcommand{\mc}{\mathcal}
\newcommand{\uu}{\;\!}

\nc{\beq}{\begin{equation}}
\nc{\eeq}{\end{equation}}
\nc{\beqa}{\begin{eqnarray}}  
\nc{\eeqa}{\end{eqnarray}}  
\nc{\bit}{\begin{itemize}}  
\nc{\eit}{\end{itemize}}  

\def\rnsig{\tau_0}

\newcommand{\eg}{{\it e.g.}}
\newcommand{\ie}{{\it i.e.}}

\usepackage{floatrow}
\newfloatcommand{capbtabbox}{table}[][\FBwidth]

\usepackage{blindtext}

\title{ 
 {\bf Dark Exoplanets}
\author{\large Yang Bai$^{\,a}$, Sida Lu$^{\,b}$, and Nicholas Orlofsky$^{\,c}$}
\date{\small \it 
$^a$Department of Physics, University of Wisconsin-Madison, Madison, WI 53706, USA\\
$^b$Institute for Advanced Study, The Hong Kong University of Science and Technology, \\Clear Water Bay, Kowloon, Hong Kong S.A.R., P. R. China \\
$^c$Department of Physics, Carleton University, Ottawa, ON K1S 5B6, Canada \\
}
}

\begin{document}

\maketitle

\setlength{\parskip}{0.2ex}

\begin{abstract}
The prevailing assumption is that all exoplanets are made of ordinary matter. However, we propose an unconventional possibility that some exoplanets could be made of dark matter, which we name ``dark exoplanets." In this paper, we explore methods to search for dark exoplanets, including the mass-radius relation, spectroscopy, missing transit, and transit light curve. Specifically, we focus on the transit light curve method and demonstrate how to distinguish partially transparent dark exoplanets from fully opaque ordinary exoplanets using both observed exoplanet data and dark exoplanet mock data. Our analysis shows that dark exoplanets with a large radius (above around 10\% of the star radius) and a small optical depth (below around one) can be identified with current telescope sensitivities.
\end{abstract}

\thispagestyle{empty}  
\newpage   
\setcounter{page}{1}  

\section{Introduction}

In the last three decades, we have seen tremendous progress on discovering new exoplanets using multi-prong detection methods including transit photometry, radial velocity (RV), direct imaging, pulsar/variable star timing, astrometry, and microlensing~\cite{Handbook}. With on-going space telescopes including Gaia~\cite{Gaia}, the Transiting Exoplanet Survey Satellite (TESS)~\cite{TESS}, the CHaracterising ExOPlanets Satellite (CHEOPS)~\cite{CHEOP} and the James Webb Space Telescope (JWST)~\cite{JWST}, along with planned future telescopes~\cite{PLATO,Ariel,Roman,TOLIMAN}, we will not only discover more exoplanets, but also study their properties in great detail. Other than searching for potentially habitable exoplanets, many science questions like planet formation and evolution, planet interiors, and atmospheric properties will be addressed. For all existing searches and studies, the exoplanets are assumed to be made of ordinary matter either explicitly or implicitly. This could be a very plausible assumption. On the other hand, there exists another possibility that some exoplanets are made not of ordinary matter, but of particles beyond the Standard Model (SM) or especially dark matter (DM) particles. We explore general search strategies for such ``dark exoplanets'' in this paper.

Dark matter is still one of the biggest mysteries of the universe, even though astrophysicist Fritz Zwicky pointed out its existence close to a century ago~\cite{Zwicky}. The DM candidates can be roughly separated into two categories: a single particle state and a composite state. The latter possibility contains ``macroscopic dark matter" with mass reaching the planet mass scale or even higher~\cite{Jacobs:2014yca,Kouvaris:2015rea,Giudice:2016zpa,Dror:2019twh,Croon:2020wpr,Bai:2020jfm,Bhoonah:2020dzs,Dhakal:2022rwn}. Dark sector models that could form macroscopic states include fermion solitons/fermi-balls \cite{Lee:1986tr,Macpherson:1994wf,Hong:2020est}, scalar nontopological solitons/Q-balls \cite{Friedberg:1976me,Coleman:1985ki,Lee:1991ax,Ponton:2019hux,Nugaev:2019vru,Heeck:2020bau,Bai:2021mzu,Bai:2022kxq,Ansari:2023cay}, dark quark nuggets \cite{Krnjaic:2014xza,Bai:2018vik,Bai:2018dxf,Liang:2016tqc}, dark nuclei \cite{Wise:2014jva,Wise:2014ola,Hardy:2014mqa,Gresham:2017zqi,Gresham:2017cvl}, and mirror sectors or any dark sector with an analog of chemistry or nuclear physics \cite{Curtin:2019lhm,Curtin:2019ngc} or degeneracy pressure \cite{Maselli:2017vfi,Hippert:2021fch,Gross:2021qgx,Ryan:2022hku}.  A macroscopic DM state with its mass and/or radius similar to those of a planet will behave as a dark exoplanet if it is bounded to a star system, even if the object's underlying physics resembles something else entirely.~\footnote{We draw a distinction between these objects made primarily of DM and other types of ``dark planets'' in previous literature that only form inside of ordinary stars \cite{Tolos:2015qra}.} In this work, we remain agnostic about the formation of stellar systems containing dark exoplanets and concentrate on how to search for a dark exoplanet (DEP) and distinguish it from an ordinary exoplanet (OEP).

As a dark-sector object, a DEP is anticipated to interact with light differently from an OEP. 
In the extreme limit without any electromagnetic interaction, methods using gravity (\eg, stellar RV, microlensing, astrometry) will be the only observation handles.
For this case, it is challenging to distinguish whether the detected exoplanet is an OEP or a DEP of the same mass. On the other hand, a DEP with a suppressed but non-negligible electromagnetic interaction is easier to be differentiated from an OEP.
For instance, different opacities of OEPs and DEPs will lead to different light curves for the transit detection method. Because other parameters are necessary to describe the planet orbit and the stellar limb darkening effects, it is still a nontrivial but achievable effort to distinguish a transiting OEP and DEP, which is the focus of this paper. 

Note that macroscopic DM is constrained by microlensing \cite{Niikura:2017zjd,Smyth:2019whb,Macho:2000nvd,EROS-2:2006ryy,OGLE,Griest:2013aaa,Oguri:2017ock}, among other constraints \cite{Bai:2020jfm}, to be less than $\mathcal{O}(10^{-2})$ of the measured DM energy density for masses greater than about $10^{23}~\text{g}$, and there are prospects to set bounds at still lower masses \cite{Katz:2018zrn,Bai:2018bej,Jung:2019fcs}.  DEPs thus need not make up all of DM, just as ordinary planets are only a small fraction of ordinary matter. Alternatively, DEPs could be all of DM if they have relatively low energy density so that they are not very massive ($<10^{23}~\text{g}$) but still have large radii, making them unobservable with present microlensing searches but potentially measurable in stellar transit searches. Interestingly, several excess microlensing events were observed by OGLE that could be interpreted as Earth-mass exoplanets \cite{Niikura:2019kqi}.

This paper is organized as follows. In Sec.~\ref{sec:detect_methods}, we discuss various ways of detecting DEPs that are orbiting stars and distinguishing them from OEPs. Secs.~\ref{sec:light_curve} and \ref{sec:analysis} explore how DEPs could appear in transit light curves, using a model-agnostic framework. Sec.~\ref{sec:light_curve} focuses on the light curve calculation and DEP model assumptions, while Sec.~\ref{sec:analysis} presents analysis of both real and mock light curves to demonstrate what sorts of transiting DEPs can be distinguished from OEPs. Sec.~\ref{sec:opacity_model} presents dark sector models for DEP interactions with light. Finally, Sec.~\ref{sec:discussion} discusses the limitations of this work and highlights our conclusions.

\section{Methods for detecting dark exoplanets orbiting stars}
\label{sec:detect_methods}

We identify several methods to detect and characterize a DEP orbiting an ordinary star. We outline the general strategy of each method in this section and expand in more detail for the transit light curve method in the next sections. 

\begin{itemize}
    \item {\it Mass-radius relation:} The masses and radii of some exoplanets can be measured independently, \eg, using RV or transit timing variation to measure the mass and transit photometry for the radius. The discovered exoplanets, with the assumption of OEPs, can be grouped into four main categories with an increasing radius $R_p$: rocky worlds, water worlds, transitional planets, and gas giants~\cite{planet-size}. An outlier exoplanet with its mass and radius not matching OEP behaviors would indicate the presence of a DEP. Some simpler cases of this could happen for a DEP candidate: the inferred planet energy density is even larger than one that is made of only Fe~\cite{2007ApJ...669.1279S}; or, an exoplanet with density less than about 0.03 g/cm$^3$. At present, no such outliers exist. If DEPs do exist, the lack of observational evidence could indicate that they have mass, radius, or weaker interactions with light outside the ability to be probed by present methods, or it could indicate they have size and density very similar to OEPs. It could also indicate that the radius has been mismeasured due to OEP model assumptions (as we will discuss for the case of transits). Alternatively, DEPs orbiting ordinary stars may be rare.

    \item {\it Spectroscopy:} During a transit, the spectrum of the sunlight can be measured and compared to the spectrum when the planet is not transiting. Differences may be indicative of absorption in the planet's atmosphere. If absorption lines can be matched to known SM absorption lines, then the planet is likely an OEP. Alternatively, an absorption spectrum that doesn't match any category of OEP may indicate that the transiting planet is a DEP. In principle, this type of analysis could also extend to light that is reflected off the exoplanet when it is not transiting, which may also be teased out by comparing to the spectrum during the secondary eclipse when the exoplanet is blocked by the star. At present there are not very many high-quality spectroscopic measurements of exoplanets, but the JWST is a promising telescope to search for DEPs~\cite{JWST_WASP39,JWST_WASP39_2}.

    \item {\it Missing transit:} A combination of RV and astrometry measurements can be used to completely determine the orbital information of an exoplanet~\cite{2014ApJ...795...41M,2017A&A...601A...9R,2019MNRAS.490.5002F,2021MNRAS.507.2856F,2021AJ....162..181L,2021AJ....162..266L,2022arXiv220804867H,2022ApJS..262...21F,2022AJ....164..196W,2023MNRAS.520.1748S}.
    If a transit is predicted based on these measurements but not detected, it could indicate the presence of a transparent DEP. In the references here, several exoplanets have the potential for transits, but the uncertainties in the measurements of the inclination angles are still too large to make definitive predictions.
    
    \item {\it Transit light curve:} Given that transit searches are the most prolific exoplanet finders, and that observations of spectroscopy or RV are currently impractical to perform on every candidate exoplanet, it is interesting to consider whether DEP candidates could be identified using transit searches alone. These candidates could then be targets for followup observations. This approach is less straightforward than the previous three approaches and is the main focus of this work.
\end{itemize}

\section{Transit light curve for partially transparent exoplanets}
\label{sec:light_curve}

Because DM must have a suppressed interaction with the photon, an exoplanet made of DM may not be completely opaque.
A DEP transit light curve will be affected as a result, and in particularly the reduction in flux should be smaller than that from a same-radius OEP. Here we present the calculation of the transit light curve of a DEP in a model-independent way.

We assume a quadratic stellar limb darkening~\cite{Kipping:2013wca} where the normalized specific surface intensity of the star surface is described by
\begin{align}\label{eq:I}
\frac{I(r)}{I(0)}=1-u_1\left(1-\sqrt{1-r^2/R^2_s}\right)-u_2\left(1-\sqrt{1-r^2/R^2_s}\right)^2\,,
\end{align}
where $r$ is the distance from the center of the star with radius $R_s$ on the image plane. The parameters $u_1,u_2$ depend on the star's properties and on the wavelength acceptance of the observing telescope. We will treat them as unknown parameters to be fitted with uniform priors, which is common in the literature, although some work has been devoted to matching these parameters to other stellar parameters~\cite{claret2017limb}.
The unblocked total ``flux'' from the star is then $\Phi_0\equiv 2\pi \int^{R_s}_0 r\uu dr\uu I(r)/I(0)=\pi R^2_s(6-2u_1-u_2)/6$, where the unit mismatch comes from the normalization factor.

\begin{figure}[t!]
\centering
    \includegraphics[width=0.65\textwidth]{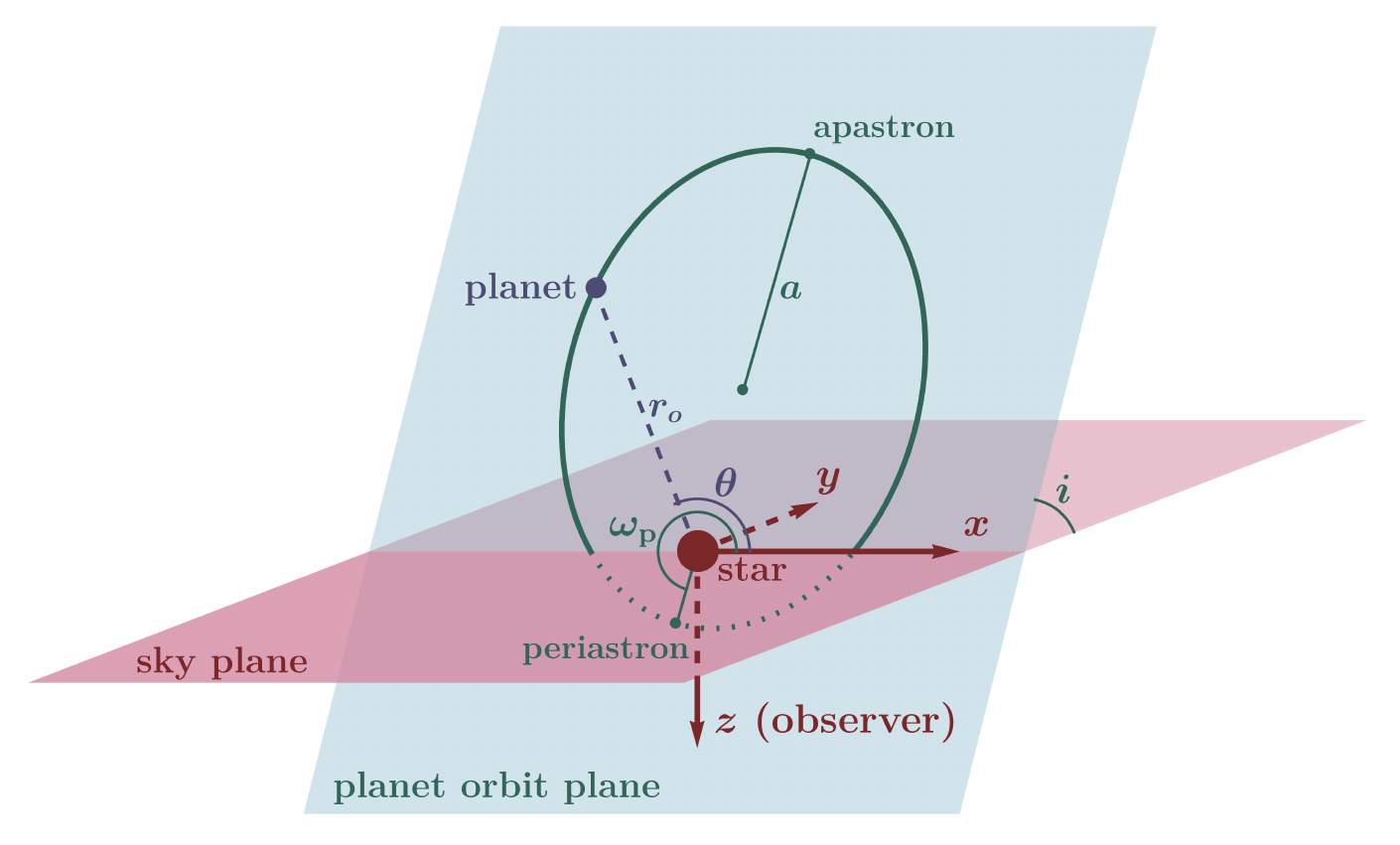}
    \caption{An illustration for the geometry of an orbiting planet. Here, $a$ is the semi-major axis; $i$ is the inclination angle; $\omega_{\rm p}$ is the argument of periapsis; $r_o$ is the distance between the planet and star centers.
    }
    \label{fig:geometry}
\end{figure}

The transmittance of an OEP is $T(x)=0$ for $x<R_p$ (the planet's radius) with $x$ the distance from the planet's center (neglecting any transparency in the atmosphere). For a DEP, it is
\begin{align}\label{eq:T}
T(x)=\exp\left(-2 \, \rnsig \sqrt{1-(x/R_p)^2}\right) \; \; \; \text{for } x<R_p \, ,
\end{align}
where $\rnsig \equiv R_p n \sigma$ denotes the characteristic optical depth. For simplicity, we assume that the dark exoplanet is made of DM particles with a constant number density $n$,~\footnote{Other number density functions make a DEP even more different from an OEP and could make the searches of DEP easier.} which is the case for some non-topological soliton DM states like Q-balls or dark quark nuggets, and the DM-photon cross section $\sigma$ as energy-independent within the telescope's energy acceptance. Further discussion of the underlying models for $\sigma$ and $n$ is delayed until Sec.~\ref{sec:opacity_model}, and for now we treat $\rnsig$ as a free, $x$-independent parameter to be fitted.

\begin{figure}[t!]
\centering
    \includegraphics[width=\textwidth]{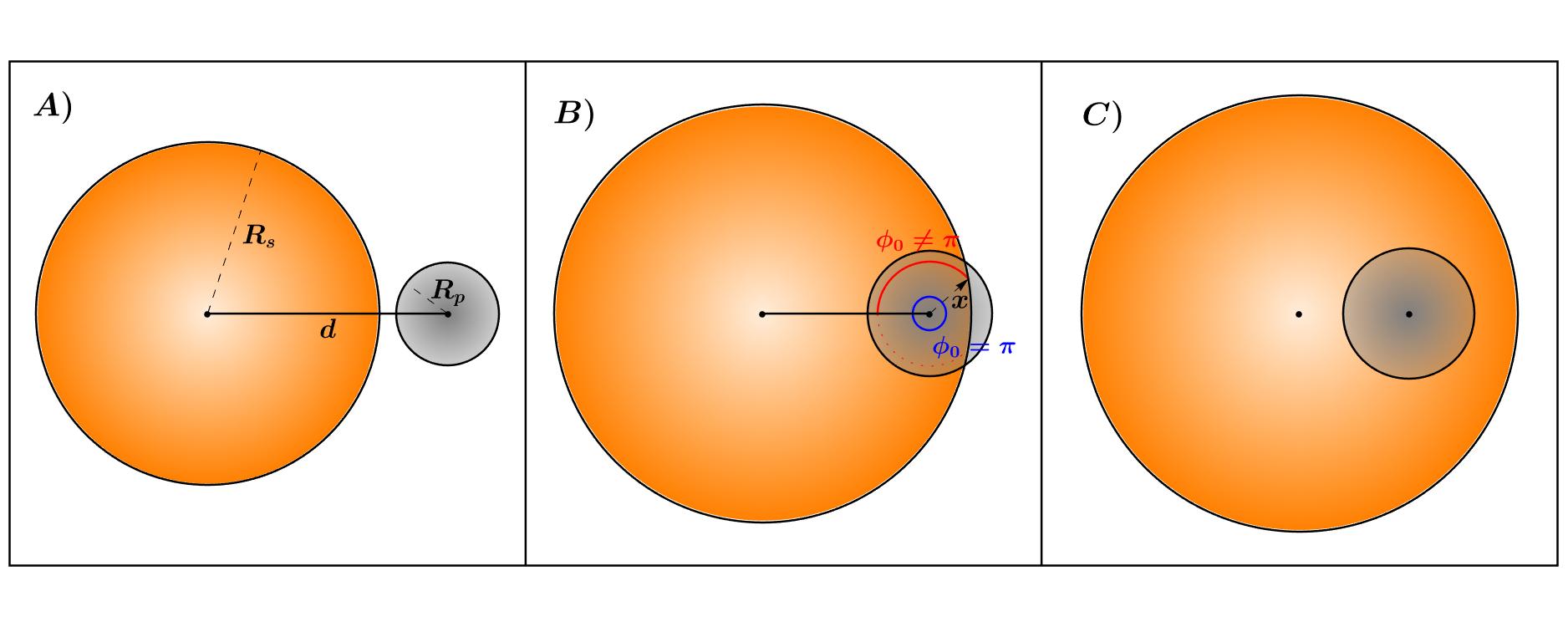}
    \caption{Some possible relative relationships between the stellar image and the DEP image. In panel A, the two images are not in contact with each other. In panel B, the DEP image partially enters the stellar image, corresponding to the ingress or egress region of the light curve. In this situation, the upper limit of the angular integral differs for different $x$, as shown in Eq.~\eqref{eq:theta0}. In panel C, the DEP image enters the stellar image completely. 
    }
    \label{fig:transit}
\end{figure}

When the exoplanet blocks the star light---\ie, when the distance $d$ between the star and planet's centers projected on the image plane satisfies $d < R_s+R_p$---the blocked flux is 
\begin{align}\label{eq:I_blocked}
\Phi_{\rm blocked}=\int^{R_p}_0 x\uu dx\uu 2\big[1-T(x)\big]\int^{\phi_0}_0 d\phi \,\frac{ I\big(\sqrt{x^2+d^2-2\uu x\uu d\cos\phi}\big)} { I(0)}\,,
\end{align}
where the definition of the quantities are given in Fig.~\ref{fig:geometry} and  Fig.~\ref{fig:transit}.
With the quadratic surface intensity given in~\eqref{eq:I}, for an OEP this integration has an analytic result~\cite{mandel2002analytic} due to the complete opaqueness of the exoplanet.
For a DEP, only the inner angular integration can be done analytically, \ie, 
\begin{align}
\int^{\phi_0}_0 d\phi \frac{ I\big(\sqrt{x^2+d^2-2\uu x\uu d\cos\phi}\big) }{ I(0) }
=& \dfrac{1}{R^2_s}\Bigg[\bigg(u_2(d^2+x^2)+(1-u_1-2u_2)R^2_s\bigg)\phi_0-2\uu u_2\uu d\uu x\sin\phi_0\nonumber\\
&  \hspace{-1.2cm} + \uu 2\uu (u_1+2u_2)\uu R_s\,\sqrt{R^2_s-(d-x)^2}\uu \uu E\left(\frac{\phi_0}{2},\frac{4\uu d\uu x}{R^2_s-(d-x)^2}\right)\Bigg] \,, 
\end{align}
where $E(\varphi,m)=\int^\varphi_0 d\phi\left(1-m\uu\sin^2 \phi\right)^{1/2}$ is the elliptical integral of the second kind.~\footnote{Numerically, it is more efficient to use Carlson elliptic integrals~\cite{Carlson} in place of the ordinary elliptic integral using the identity $E(\varphi,m)=\sin\varphi\uu R_F(\cos^2\varphi,1-m\uu\sin^2\varphi,1)-\frac{1}{3}m\sin^3\varphi\uu R_D(\cos^2\varphi,1-m\uu\sin^2\varphi,1)$ with the two functions $R_F$ and $R_D$ defined in Ref.~\cite{Carlson}.}
Depending on the relative location between the stellar image and the dark exoplanet image, the angular integration upper limit is given by
\begin{align}\label{eq:theta0}
\phi_0=\begin{cases}
\arccos{\left(\dfrac{x^2+d^2-R^2_s}{2\uu x\uu d}\right)} \, , & x>R_s-d \, , \\
\pi \, , & x \leqslant  R_s-d \, .
\end{cases}
\end{align}
The normalized light curve is then calculated by $(\Phi_0-\Phi_{\rm blocked})/\Phi_0$.

\section{Data analysis of transit light curves}
\label{sec:analysis}

As a proof of principle, we fit both ordinary and dark exoplanet transit light curves to both real and simulated phase-folded data of exoplanet transits, assuming a one-star--one-planet system with a circular orbit. For simplicity, we take the period $T_p$ as sufficiently well measured that we do not refit it (for real data, we use published values, and for simulated data we use the known simulated period). The mass $M_s$ and radius $R_s$ of the host star are also assumed to be known from independent measurements and modeling, see \cite{Handbook} for a review. Hence the semi-major axis of the exoplanet orbit $a$ can be determined. For an ordinary exoplanet, the remaining planetary parameters to fit are the planet radius $r_p \equiv R_p/R_s$, orbital inclination angle $i$, quadratic limb darkening coefficients $u_1$ and $u_2$, and orbital angular position of the first data point on the phase-folded light curve $\theta_{\rm init}$ necessary for aligning the transit centers in the model and data (see Fig.~\ref{fig:geometry} for the definition of $\theta$). 
For a dark exoplanet, we must also fit $\rnsig$ so that $T(x)$ in Eq.~(\ref{eq:T}) can be determined. Altogether, we have 5 and 6 model parameters for the OEP and DEP models, respectively.

If the transit data from a partially opaque DEP is fitted assuming a fully opaque OEP, then the OEP fit is expected to give a smaller planetary radius than the DEP one to match the observed maximum reduction in flux. Then, to keep the same time interval between the end of the ingress and start of the egress, the inclination angle $i$ should move further away from $\pi/2$. Even with these adjustments, the ingress and egress may not be well-fitted by an OEP, which may be partially ameliorated by the limb-darkening coefficients or by changing the orbital eccentricity. If the transit does not have clearly defined ingress and egress regions, it may be more difficult to distinguish DEPs from OEPs.

As we will demonstrate with both real and simulated data, it is easier to distinguish between OEPs and DEPs when the planet radius is large. Intuitively, this is because the transmittance function $T(x)$ becomes better resolved for a larger exoplanet. Additionally, a DEP must be sufficiently transparent, satisfying $\rnsig \lesssim \mathcal{O}(1)$, to be distinguishable from an OEP. Finally, transits that only graze the edge of the star are more difficult to distinguish because they contain less information about the exoplanet opacity profile.

\subsection{DEP searches with confirmed exoplanet light curves}

Here, we analyze data from two transiting exoplanets, which could serve as examples for examining more transit light curve data. These particular data were chosen to illustrate how constraining real data can be on the DEP hypothesis, rather than the exoplanets that would be ``most likely'' to be DEPs, and a full analysis of all the transit data on all known/candidate exoplanets is well beyond the scope of this work. Unsurprisingly, the data analyzed in this section favors the OEP hypothesis over the DEP hypothesis. However, the transit data do not convincingly rule out the DEP hypothesis when taken on their own. Moreover, differences in the type of data and type of exoplanet being observed lead to differences in how convincingly the DEP hypothesis can be excluded.

\begin{table}[t!]
    \centering
    \renewcommand{\arraystretch}{1.1}
    \addtolength{\tabcolsep}{5pt} 
    \begin{tabular}{c|c|c|c|c}
        \hline\hline
         & \multicolumn{2}{c|}{CoRoT-1\,b} & \multicolumn{2}{c}{K2-44\,b}\\
        \hline\hline
        $a\,(R_s)$ & \multicolumn{2}{c|}{4.910} & \multicolumn{2}{c}{8.760}  \\
        $T_p$ (days) & \multicolumn{2}{c|}{1.509} & \multicolumn{2}{c}{5.655}  \\
        \hline\hline
         & OEP & DEP & OEP & DEP\\
        \hline
        $i\,({\rm rad})$ & 1.486 & 1.486 & 1.525 & 1.525  \\
        $\theta_{\rm init}\,({\rm rad})$ & 4.433 & 4.433 & 4.041 & 4.041\\
        $u_1$ & 0 & 0 & 0.618 & 0.619 \\
        $u_2$ & 0.705 & 0.705 & 0.228 & 0.227 \\
        $r_p$ & 0.136 & 0.137 & 0.0163 & 0.0163 \\
        $\rnsig$ & --- & 81.46 & --- & 59.23 \\
        \hline
        $\chi^2$ & 127.97 & 127.95 & 858.10 & 858.09\\ 
        degrees of freedom & 97  & 96 & 659 & 658 \\
        \hline \hline
    \end{tabular}
    \caption{The best-fit parameters of CoRoT-1\,b and K2-44\,b for both OEP and DEP models. The orbits are assumed to be circular for simplicity. The orbital radii and periods are fixed to be the same values as the accepted ordinary exoplanet measurements~\cite{Barge:2008ts,del2020tfaw}, as the period is very well measured. The orbital inclination $i$, initial phase of the folded light curve $\theta_{\rm init}$, limb darkening coefficients $(u_1, u_2)$, and exoplanet radius $r_p \equiv R_p/R_s$ are fitted in both cases, while the dark exoplanet case has an additional fitted parameter of the optical depth $\rnsig$.
    }
    \label{tab:BM_regular}
\end{table}

The target exoplanets we use for analysis are CoRoT-1\,b~\cite{Barge:2008ts} and K2-44\,b~\cite{del2020tfaw}, which are chosen to demonstrate the two extremes of large- and small-radius planets in one-star--one-planet systems with nearly circular orbits. For CoRoT-1\,b, we use the processed light curve data from the VLT observatory in~\cite{gillon2009vlt}.
For K2-44\,b, we use the pixel level decorrelated, cotrending basis vector (CBV) corrected data provided by the {\tt EVEREST 2.0} pipeline~\cite{luger2018update}.
Similar to the treatment in~\cite{del2020tfaw}, we use the first 3072 epochs out of those with {\tt QUALITY=0} flag to generate the light curve, and produce the normalized folded light curve through {\tt EVEREST} with parameters $T_0=1978.7248$ (BJD-2454833), period = 5.6549\,days, and duration = 0.3\,days (though see \cite{del2020tfaw} for possible improvements to the data processing).

\begin{figure}[t!]
\centering
    \includegraphics[width=0.45 \textwidth]{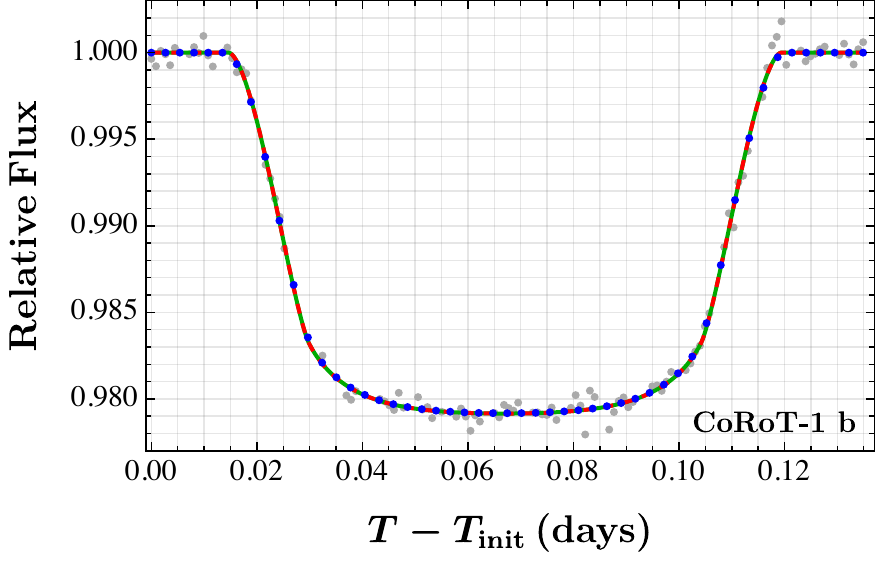}
    \hspace{8mm}
    \includegraphics[width=0.45 \textwidth]{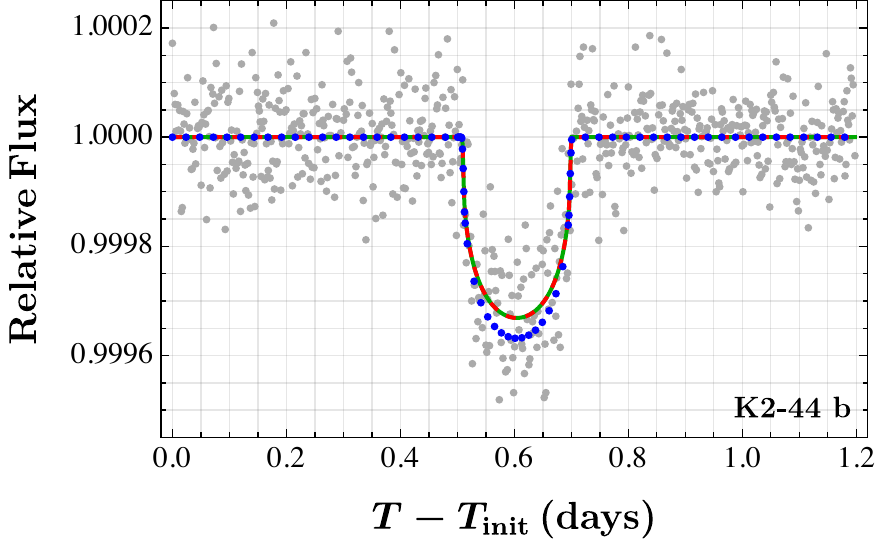}
    \caption{The measured (normalized and folded) transit light curves of CoRoT-1\,b (left) and K2-44\,b (right) together with light curves from both ordinary and dark exoplanet model fits. $T_{\rm init}$ indicates the time of the leftmost measurement in the panel after phase folding the light curves. The solid green and dashed red curves represent the best-fit ordinary and dark exoplanet model parameters as given in Table~\ref{tab:BM_regular}, and the dotted blue curves correspond to the blue dots in Fig.~\ref{fig:chi2_contours}.
    }
    \label{fig:light_curves}
\end{figure}

\begin{figure}[t!]
\centering
    \includegraphics[width=0.434 \textwidth]{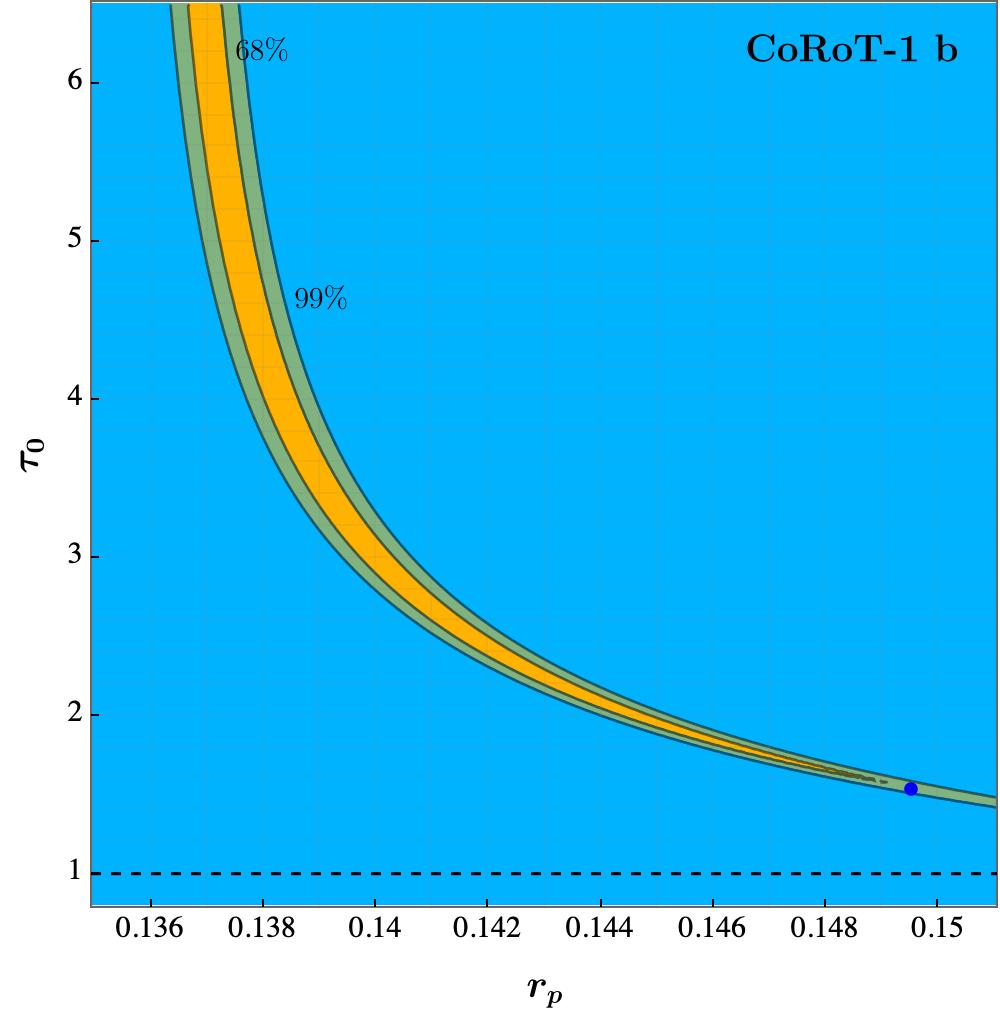}
    \hspace{0.6cm}
    \includegraphics[width=0.46 \textwidth]{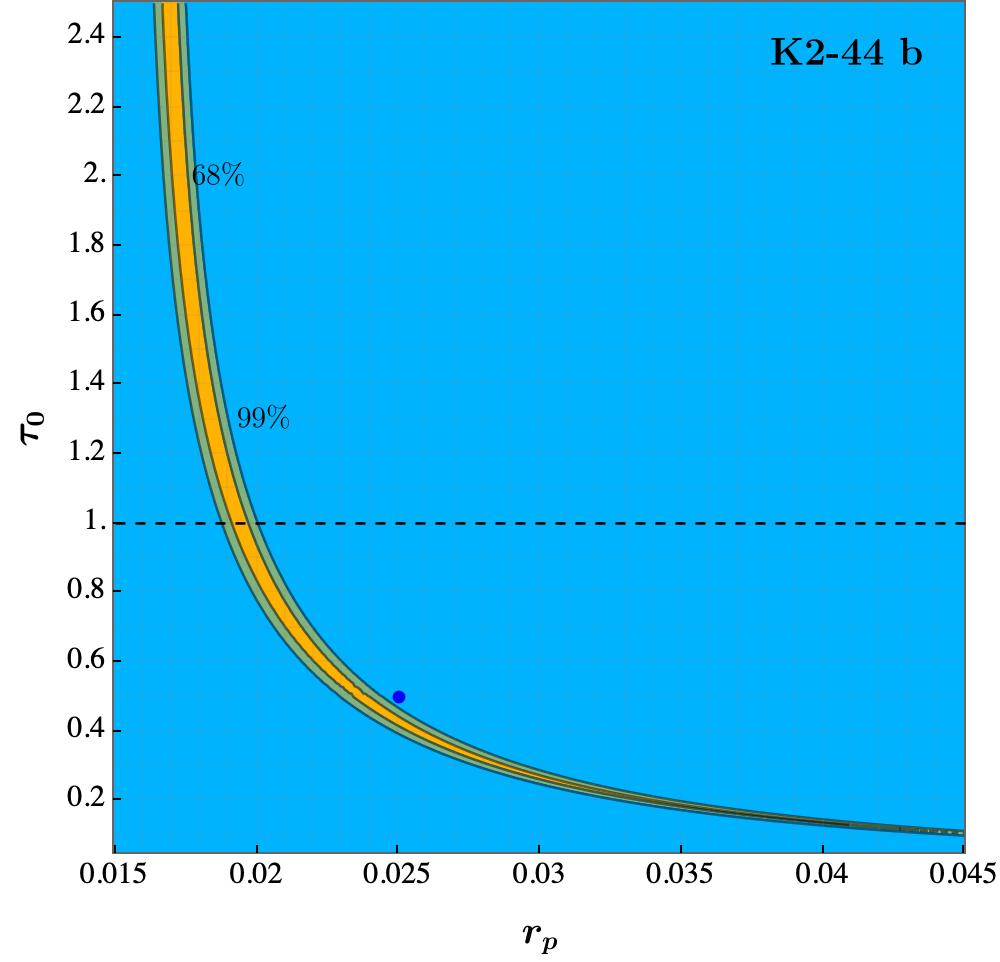}
    \caption{Contours of $\Delta\chi^2$ from the best-fit point of DEP for the CoRoT-1\,b and K2-44\,b measurements. All parameters except $r_p$ and $\rnsig$ are fixed (instead of marginalized) to be the same as the best-fit point in Table~\ref{tab:BM_regular}. The yellow and green contours differ from the best-fit point at 68\% and 99\% confidential level for two degrees of freedom. The blue point in each panel corresponds to the blue dotted light curve in the corresponding panel of Fig.~\ref{fig:light_curves}.
    }
    \label{fig:chi2_contours}
\end{figure}

The best-fit OEP and DEP parameters are given in Table~\ref{tab:BM_regular}.~\footnote{Despite the similarity in data selection and processing, our data used for the fit are not completely the same as those in~\cite{del2020tfaw}, such that our best-fit parameters for the OEP case are slightly different from the currently accepted values.}
Using a chi-square difference test, we can determine if the goodness of fit is statistically improved by including the optical depth parameter in the DEP model. If so, then the data may be able to distinguish between an OEP and DEP; otherwise no evidence is found to reject the hypothesis of an OEP. 
The $\chi^2$ difference does not favor the DEP hypothesis over the OEP hypothesis. Further, the DEP best-fit point has a value of $\rnsig \gg 1$ such that it is very opaque and essentially indistinguishable from an OEP anyways. The corresponding light curves are displayed in Fig.~\ref{fig:light_curves}, in which the best-fit light curves from the two models almost completely overlap with each other, a well-expected result as the minimal $\chi^2$ of the two models are almost the same.
The dotted blue curves in Fig.~\ref{fig:light_curves} correspond to the blue dots in Fig.~\ref{fig:chi2_contours}, which serve as illustrations of deviating from the best-fit region.

On the other hand, while the DEP hypothesis is not favored, it cannot always be excluded. To demonstrate this point, in Fig.~\ref{fig:chi2_contours} we make contour plots of the chi-square goodness of fit on $r_p \equiv R_p/R_s$ and $\rnsig$, keeping the other parameters fixed to their best-fit value.~\footnote{The credible regions are calculated using a $\chi^2$ difference test with two degrees of freedom. In principle, the other parameters could be marginalized to their best-fitting value at each point on this 2D plane, \eg, using a Markov Chain Monte Carlo to sample the posterior distribution. However, the result would not differ much because most of the parameters would still be very close to their best-fit value, as demonstrated by Table~\ref{tab:BM_regular}. The credible regions may be slightly enlarged, but their qualitative features would remain.} 
The two parameters are inversely correlated. 
At smaller $r_p$ the contours tend to be insensitive to $\rnsig$, as the OEP model can be effectively viewed as a DEP model in the $\rnsig\to\infty$ limit.
At larger $r_p$ the contours of fixed likelihood pinch off, setting a lower bound on $\rnsig$. The 68\% confidence bound for CoRoT-1~b is $r_p < 0.149$ and $\rnsig > 1.6$, while for K2-44~b it is $r_p < 0.044$ and $\rnsig > 0.1$. 
The $\tau_0$ bound is stronger for CoRoT-1\,b than for K2-44\,b, which can be attributed to CoRoT-1\,b's larger radius. 
Note that in the limit of small $\tau_0$ and large radius, the best-fit region follows the relationship $n \sigma \propto r_p^{-3}$ (or $\tau_0 \propto r_p^{-2}$), where two factors of $r_p^{-1}$ come from the cross-sectional area of the dark exoplanet, one factor of $r_p^{-1}$ comes from the opacity [where $1- T(x) \approx 2 \rnsig (1-x^2/R_p^2)^{1/2} + \mathcal{O}\big(\rnsig^2\big)$ for small $\rnsig$].

It may be noted that while we have conservatively allowed the values of the limb darkening parameters to float to their best-fit points (subject to physical constraints on the brightness being monotonically decreasing from the center and nonnegative), there has been work on classifying the limb darkening parameters based on other stellar properties \cite{claret2017limb}. If we impose these empirical constraints as priors on our fits, they may change the results on a case-by-case basis.

\subsection{Mock light curves of dark exoplanets}
\label{sec:mock}

In addition to reexamining the measured transit light curves and searching for DEPs, it is also worthwhile to check whether a mock DEP transit light curve can always be well mimicked by an OEP. 
Parameter space where the OEP fits fail to explain the data would be the natural target region for DEP hunting.

The mock data are generated in the following way.
The mass of the host star $M_s$ is uniformly chosen from $0.5-5M_\odot$, where $_\odot$ represents the Sun (note that $\gtrsim 90\%$ of the exoplanets confirmed with the transit method have their host star heavier than $0.5M_\odot$).
The stellar radius is determined as $R_s=e^g\uu R(M_s)$, with
\begin{align} 
    \dfrac{R(M_s)}{R_\odot}=\begin{cases}
        0.89\left(M_s/M_\odot\right)^{0.89} \uu, & M_s\leqslant 1.66M_\odot \uu, \vspace{1mm}\\
        1.01\left(M_s/M_\odot\right)^{0.57}\uu,  & M_s>1.66M_\odot \uu,
    \end{cases}
\end{align}
which is the fitted mass-radius relation for ZAMS models~\cite{demircan1991stellar}, and $g$ a random variable from a normal distribution $\mc{N}(\mu=0, \sigma=0.5)$ responsible for the deviation from this relation.~\footnote{Actually, the stellar radii for confirmed exoplanets are slightly larger than the ZAMS model, likely due to selection bias \cite{Gaidos_2012,2022arXiv220703019P}, but the effect is small enough that we neglect it here. We focus here on main sequence stars because almost all exoplanets discovered using the transit method orbit main sequence stars~\cite{Handbook}. While only a few exoplanets are known around white dwarfs \cite{2021orel.bookE...1V}, transits of smaller objects like minor planets (radius $\lesssim 1000~\text{km}$) are easier to discern in white dwarf systems, which could make them interesting candidates for future study of smaller DEPs.}
The orbit of the DEP is assumed to be circular, and the period is uniformly chosen from 1-20 days, a region where $\sim 75\%$ of the current measured transit events lie (due in part to observational bias).
The semi-major axis $a$ is then determined with the Kepler's third law, and we require $a/R_s>2$.
The limb darkening coefficients $u_1$ and $u_2$ are uniformly chosen on the region $u_1>0,\, u_1+u_2<1$ and $u_1+2u_2>0$, which enforces that $I(r)$ is monotonically decreasing and non-negative. 
The radius of the DEP is uniformly chosen from $r_p \in [0.01,0.15]$, and the orbital plane inclination angle $i$ is uniformly chosen between $\arccos[(R_p+R_s)/a]$ (such that there will be a transit event) and $\pi/2$.
The optical depth parameter is chosen uniformly from $\rnsig \in [0.1,3]$.
Each simulated light curve is chosen to cover an orbital angular position $\theta \in \big[(1.5-0.01n_t)\pi,~(1.5+0.01n_t)\pi\big]$, where $n_t$ is the minimal integer such that $\theta=(1.5-0.01n_t)\pi$ is earlier than the corresponding DEP's ingress (note that for a circular orbit the transit is symmetric around $\theta=1.5\pi$). This angular position region in $\theta$ is evenly divided into $N$ segments, where $N$ is randomly uniformly chosen on $[100, 250]$, and on each segment an angular position is chosen {\it randomly} and uniformly, as the light curve used for fitting is phase-folded. The error bar size of data point $j$ is set to be $e_j=\sigma_e+\mc{N}(0, \Delta\sigma_e)$, where we choose $\sigma_e=10^{-4}$ and $\Delta\sigma_e=10^{-6}$, values comparable to the error bar sizes and fluctuations of the current measurements from Kepler and TESS~\cite{TESS}.
The center value of this data point is then set to be its model value from Sec.~\ref{sec:light_curve} plus a random number from $\mc{N}(0, e_j)$.

A total of 1280 sets of mock light curves are generated with the method above, and each of them are fitted with both the DEP and the OEP models.
The corresponding minimal $\chi^2$ for the best-fit points are denoted as $\chi^2_{\rm D}$ and $\chi^2_{\rm O}$, respectively.
In addition, we calculate the $\chi^2$ value corresponding to a null hypothesis (\ie, constant relative flux $\equiv 1$), and denote it as $\chi^2_{\rm null}$.
We identify a mock light curve $k$ with $N_k$ data points as a DEP candidate if it meets the following three criteria: (i) the null-hypothesis is rejected at 99\% confidence level, (ii) the DEP model is accepted at 95\% confidence level and (iii) the OEP model is rejected at $\alpha = 1\%$ or 5\% significance level.
In other words, letting $F_n(\chi^2)$ be the cumulative distribution function for a chi-square distribution with $n$ degrees of freedom, we require $F_{N_k-1}(\chi^2_{\rm null})>0.99$, $F_{N_k-6-1}(\chi^2_{\rm D})<0.95$ and $F_{N_k-5-1}(\chi^2_{\rm O})>1-\alpha$.
The light curves that satisfy (i) but fail (iii) with $\alpha=5\%$ are identified as OEP candidates, regardless if they satisfy (ii). The rest (not in DEP or OEP categories) are classified as undetectable.
A total of 120 sets of mock light curves are classified as undetectable in this way.
Many of the undetectable light curves correspond to grazing inclination angles $i$ (near their minimum for given $a$) or small $r_p$, both of which are harder to detect. Note that the criteria for OEP vs DEP candidates is ``conservative'' in the sense that if an OEP model provides a good fit, the transiting planet is not classified as a DEP candidate even if the DEP model significantly improves the goodness of fit compared to the OEP.

The most important parameters that determine whether a DEP can be distinguished from an OEP are $r_p$ and $\rnsig$. As demonstrated in the left panel of Fig.~\ref{fig:scatter_rp_rnsig} (for a full corner plot, see Appendix~\ref{sec:appendix}), DEPs with large $r_p \gtrsim \mathcal{O}(0.1)$ and small $\rnsig \lesssim \mathcal{O}(1)$ are the most easily distinguished.
This result can be understood in the following way.
The DEP light curves distinguish themselves from those of OEPs mainly through the shape of the ingress/egress region, see the bottom-right panel of Fig.~\ref{fig:scatter_rp_rnsig} for the fitted light curves in one of the DEP candidates.
A larger radius and more transparent (smaller $\rnsig$) DEP has a better chance to manifest such differences.

\begin{figure}[t!]
\centering
    \includegraphics[width=0.48\textwidth]{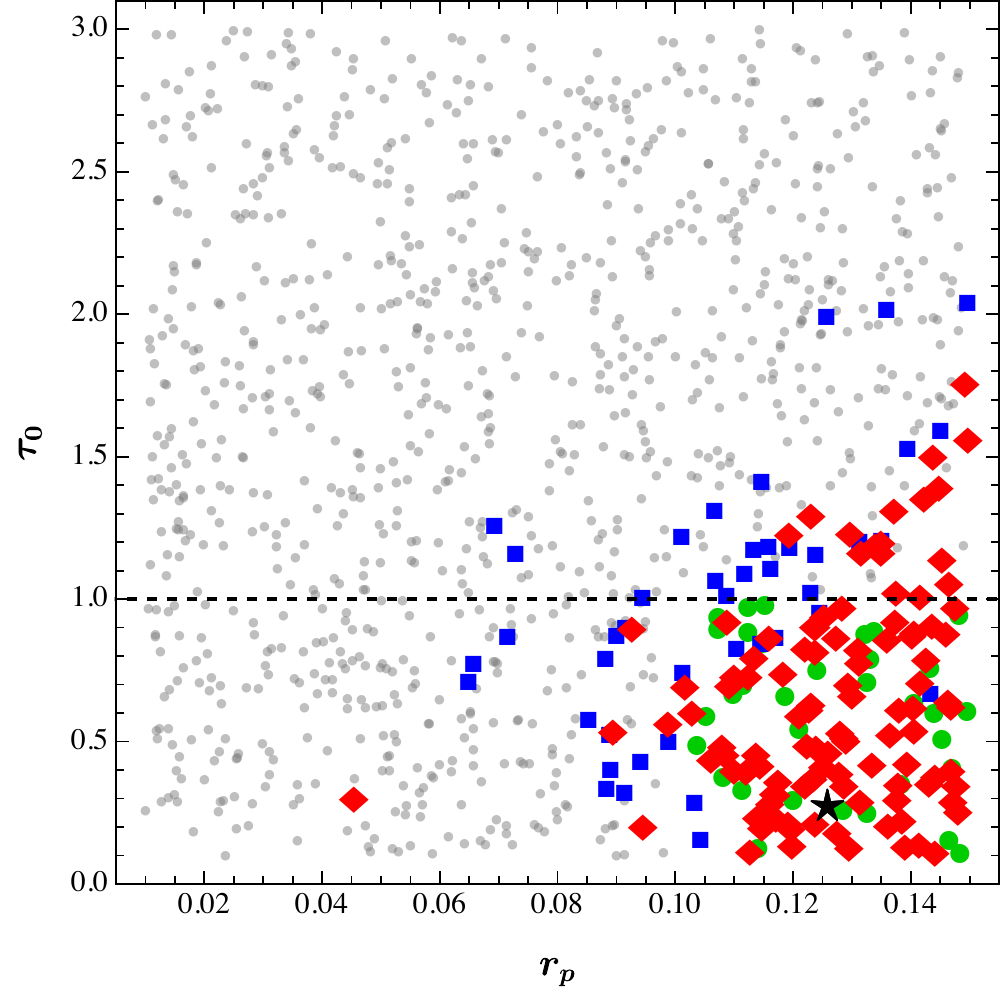}
    \hspace{1mm}
    \includegraphics[width=0.495\textwidth]{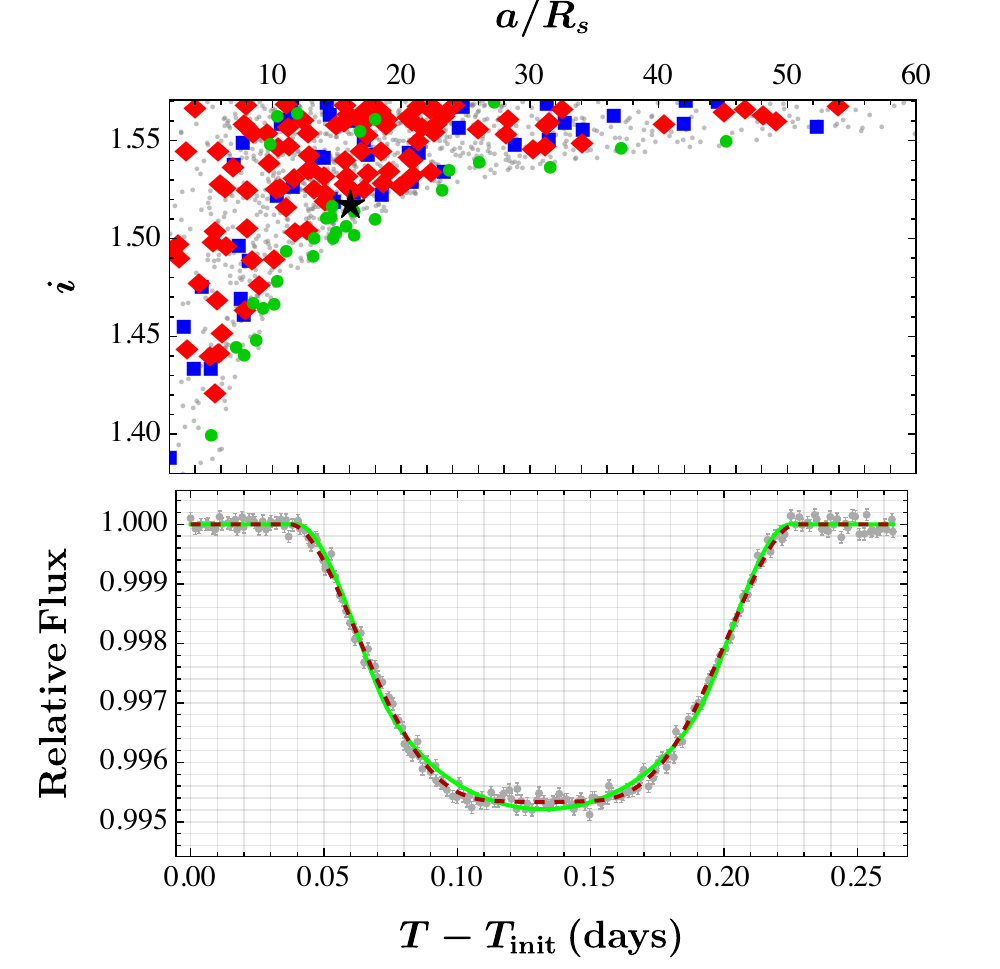}
    \caption{
    {\it Left:} The scatter plot of all DEP and OEP candidates on the true-value $r_p-\rnsig$ plane. The red diamonds (and black star) are the identified DEP candidates when the OEP model is rejected at $1-\alpha=99\%$ confidence level, and the blue squares are at $1-\alpha=95\%$ confidence level. 
    Both the small gray and big green dots are OEP candidates.
    The green dots are highlighted as they have $r_p>0.1$ and $\rnsig<1$, a similar parameter region to most of the DEP candidates.
    Mock light curves that are classified as undetectable are not shown. 
    {\it Top-Right:} The same mock data on the true-value $a-i$ plane, with the point style the same as the left panel.
    {\it Bottom-Right:} The best-fitted DEP (dashed red) and OEP (solid green) light curves for the DEP candidate indicated by the black star in the other two panels. 
    }
    \label{fig:scatter_rp_rnsig}
\end{figure}

There are still DEP light curves with small $\rnsig$ and large $r_p$ that cannot be distinguished from OEPs. In Fig.~\ref{fig:scatter_rp_rnsig}, the OEP candidates for light curves generated with $r_p>0.1$ and $\rnsig <1$ are highlighted as green dots. 
Almost all of them have inclination angles $i$ near their $a$-dependent minima as shown in the top-right panel, meaning that these mock exoplanets just graze the outer edge of their stellar images. Thus, less information is present to distinguish between the stellar limb darkening and DEP opacity effects. On the other hand, most of the red points (and blue points, to a lesser extent) have inclination angle closer to $\pi/2$. This is because a transit trajectory closer to the stellar center has a longer transit time and ``scans'' the stellar surface brightness and DEP opacity profiles more thoroughly, making the difference between the two types of light curves more manifest.

The bottom-right panel of Fig.~\ref{fig:scatter_rp_rnsig} shows one example light curve from the mock data where the OEP hypothesis is rejected with $\alpha = 0.01$ but the DEP hypothesis is acceptable. This mock data point is denoted by the black star in the other two panels. The light curves from the best-fitted DEP model and the OEP model are shown in the dashed-red and green lines, respectively.  Note that the best-fit OEP light curve does not show good agreement, particularly near the edges of the ingress and egress regions. The mock data contains 201 data points, and has a true value of $(i, r_p, u_1, u_2, \rnsig, \theta_{\rm init})=(1.517, 0.126, 0.007, 0.151, 0.274, 4.650)$. The OEP model gives a best fit of $(i, r_p, u_1, u_2, \theta_{\rm init})=(1.521, 0.073, 0.869, 0.131, 4.650)$. In particular, the OEP model best fit has a much smaller radius $r_p$ than the true value to compensate for its opacity and a bit larger inclination angle $i$ to keep the transit duration fixed with decreasing~$r_p$. The OEP model fit for this light curve also prefers $I(R_s)/I(0) = 1 - u_1 - u_2 \approx 0$, whereas the true value is closer to 0.84. This could be because this particular transit has a somewhat grazing transit with $i$ near its $a$-dependent minimum (see the top right panel of Fig.~\ref{fig:scatter_rp_rnsig}). Therefore, the small DEP opacity can be partially mimicked by an OEP orbiting a star with dim limbs---either results in a smaller fractional darkening. In fact, as discussed in Appendix~\ref{sec:appendix}, it may be easier to distinguish DEPs orbiting stars with larger true values of $I(R_s)/I(0)$, as is the case for this example mock data point.

\begin{figure}[t!]
\centering
    \includegraphics[width=0.5\textwidth]{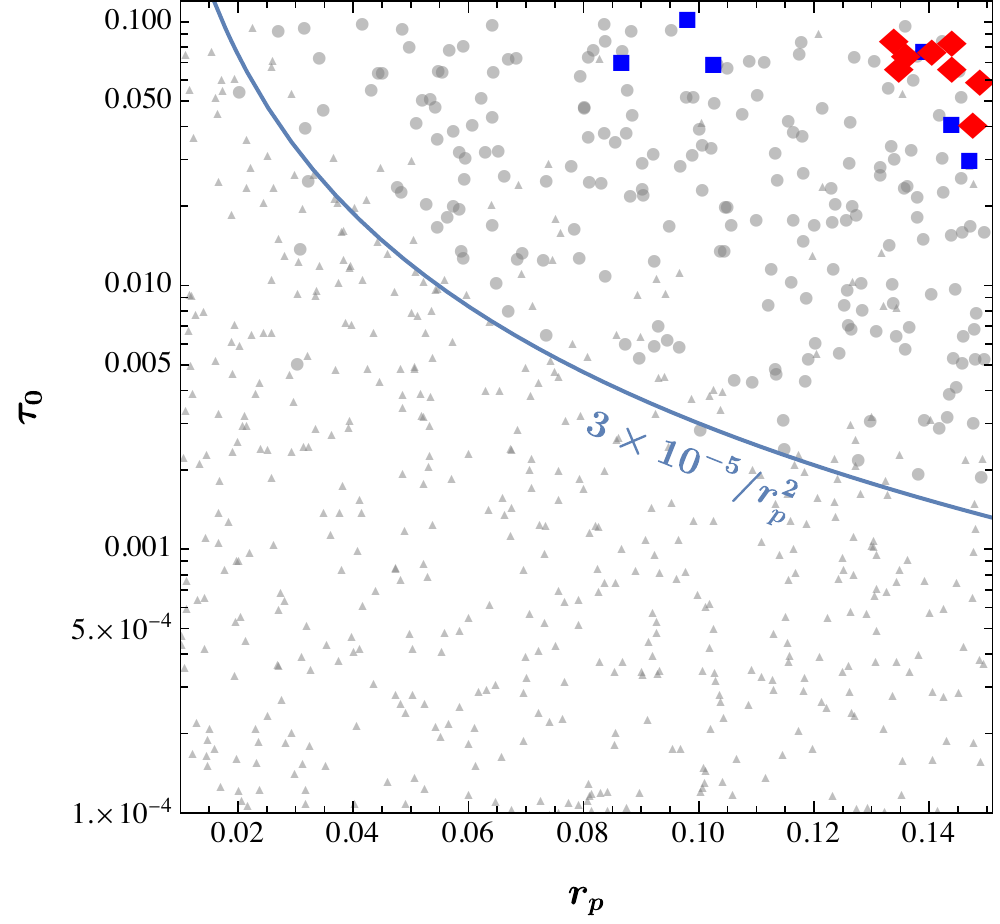}
    \caption{Similar to the left panel of Fig.~\ref{fig:scatter_rp_rnsig}, but with another set of mock data focusing on small $\rnsig<10^{-1}$. Light brown triangles indicate mock DEPs that are classified as undetectable. The gray circle points are selected as OEP candidates. All other point styles are the same as Fig.~\ref{fig:scatter_rp_rnsig}. The blue line indicates $r_p^2 \rnsig = 3 \times 10^{-5}$. 
    }
    \label{fig:scatter_rp_small_rnsig}
\end{figure}

To investigate how transparent a DEP can be before it is undetectable, we simulate another 1000 mock DEP light curves sampling logarithmically in $\rnsig \in [10^{-5},10^{-1}]$ but with all other sampling criteria the same as before. The fitted results are plotted in Fig.~\ref{fig:scatter_rp_small_rnsig} in analogy to the left panel of Fig.~\ref{fig:scatter_rp_rnsig}. All of the OEP candidate points are shown in gray circle points here. The light brown triangles indicate those where the DEP mock light curves are classified as undetectable, \ie, indistinguishable from a flat light curve. We expected that points with $r_p^2 \rnsig \lesssim \sigma_e / \sqrt{N_\text{data}} \sim 10^{-5}$ would be undetectable, and indeed the boundary between detectable and undetectable exoplanets are roughly described by the blue curve of $r^2_p\tau_0=3\times 10^{-5}$, where the factor 3 is chosen manually. However, a much larger $r_p^2 \rnsig$ than this is necessary to differentiate between OEPs and DEPs. Specifically, the DEP candidates satisfy $r_p \gtrsim 0.1$ (as before) and $\rnsig \gtrsim 0.05$. This is because for too small $r_p^2 \rnsig$, the reduction in flux in the transit light curve is too shallow to distinguish the two models.

\section{Model-dependent opacity of dark exoplanets}
\label{sec:opacity_model}

In the previous sections, the dark exoplanet is modeled to be composed of dark particles with a constant number density $n$ and cross section to interact with light in the wavelengths being observed $\sigma$ (which may in principle depend on wavelength, especially if the dark sector is complex and contains its own dark chemistry, but which we take as constant here for visible wavelengths measured by transit-hunting telescopes). For instance, dark quark nuggets~\cite{Bai:2018dxf} or non-topological solitons like Q-balls~\cite{Coleman:1985ki} contain either fermion or complex scalar fields as their constituents, and mirror sectors may contain atom-like states that could form dark planets (or stars~\cite{Curtin:2019lhm,Curtin:2019ngc}). Here, we elaborate further on one DEP model based on physics similar to degenerate compact stars.

Similar to white dwarfs or neutron stars, where the degenerate Fermi pressure balances the gravitational pressure, the dark exoplanet could contain constituent particles with neutral, positively charged, and negatively charged particles with masses $m_0, m_+, m_-$ (see also \cite{Kouvaris:2015rea,Giudice:2016zpa,Maselli:2017vfi,Hippert:2021fch,Gross:2021qgx,Ryan:2022hku}). Note that $m_+$ is not necessarily equal to $m_-$, similar to the unequal mass of electron and proton inside a hydrogen atom. In principle, cosmologically stable and abundant electromagnetically-charged dark-sector particles are severely constrained by astronomical and cosmological observations, \eg, the cosmic microwave background~\cite{Gluscevic:2017ywp} and the heating of galactic clouds~\cite{Bhoonah:2018wmw}.
These constraints can be relieved if the dark particles are bound into neutral composites by dark interactions before the gravitational interaction becomes important, or if these charged states are only responsible for a fraction of the observed dark matter abundance~\cite{Berlin:2018sjs} (see Refs.~\cite{Appelquist:2015zfa,Kribs:2016cew} for models with strong dynamics in the dark sector and a neutral composite dark baryon state coupling to photon at higher moments). Alternatively, the heavier charged states may decay into lighter neutral states and charged SM particles in analogy to the neutralino and chargino in supersymmetry (or the neutron and proton if their masses were reversed), and the charged states may only be populated inside of DEPs where the degenerate Fermi energy is sufficiently high and/or there is a deconfined vacuum state where the charged states have smaller masses.  
Similar to the case of hydrogen where the electron and proton carry individual global symmetries resulting in stable atoms, the annihilations of positively and negatively charged particles are forbidden by some dark flavor symmetries. For simplicity, we assume that the electrically charged particles are much heavier than the neutral particle, $m_+ \approx m_- \gg m_0$, and an equal number density for the three species. Similar to the white dwarf case where the lighter fermion (electron) provides degenerate Fermi pressures while the heavier fermion (proton) provides the main mass, the DEP has the following radius and mass relation
\beqa
R_p \approx \frac{3^{4/3}\,\pi^{2/3}\,M_{\rm pl}^2}{8\,m_0\,m_+^{5/3}\,M_p^{1/3}} 
= 1.7\,R_\oplus\times \left(\frac{10\,\mbox{keV}}{m_0}\right)\,\left(\frac{100\,\mbox{GeV}}{m_+}\right)^{5/3}\,\left(\frac{M_\oplus}{M_p}\right)^{1/3}
~,
\eeqa
in the non-relativistic limit with Fermi momentum $p_F\approx\frac{4}{3^{2/3}\,\pi^{1/3}}\,m_0\,m_+^{4/3}\,M_p^{2/3}\,M_{\rm pl}^{-2}$ and $M_\text{pl}=1/\sqrt{G}=1.22 \times 10^{19}~\text{GeV}$. For the nonrelativistic approximation to be valid, $m_+ \ll M_p^{-1/2}\,M_{\rm pl}^{3/2} = 736\,\mbox{GeV}\times(M_\oplus/M_p)^{1/2}$. Here, $M_\oplus$ is Earth's mass.

The optical depth is determined by integrating the dark exoplanet density along the line of sight of the light,
\beqa
\tau &=& \int d \ell \, (n_+ + n_-) \, \sigma_{\rm T}  \sim 2\,n_+ \,\sigma_{\rm T} \,R_p = \frac{128\,\alpha^2\,m_0^2\,m_+^{1/3}\,M_p^{5/3}}{3^{8/3}\,\pi^{4/3}\,M_{\rm pl}^4}
\nonumber \\
& =& 1.2 \times 10^{-4}\times \left(\frac{m_0}{10\,\mbox{keV}}\right)^2\,\left(\frac{m_+}{100\,\mbox{GeV}}\right)^{1/3}\,\left(\frac{M_p}{M_\oplus}\right)^{5/3} \, ,
\label{eq:opacity1}
\eeqa
where the ``Thomson cross section" for photon scattering off a heavy charged particle with $\sigma_{\rm T} = 8\pi\,\alpha^2/(3\,m_+^2)$ is used. 

For a dark exoplanet model different from a white dwarf, one could have other pressures like the pressures from dark particle self interactions~\cite{Lee:1991ax} or thermal pressure~\cite{Gross:2021qgx}, with or without the vacuum pressure playing a role. The relations of $R_p$ and $M_p$ with charged-particle masses could depend on other model parameters. Therefore, one could still treat $R_p$ and $M_p$ as independent parameters to discuss the opacity of a DEP.   
If electrically charged particles contribute a sizable fraction to the total DEP mass, the charged-particle number density is $\mathcal{O}[M_p/(2 m_+ \,R_p^3)]$ and the optical depth is
\beqa
\tau &=& \int d \ell \, (n_+ + n_-) \, \sigma_{\rm T}  \sim 2 n_+ \,\sigma_{\rm T} \,R_p = \frac{2\,\alpha^2\,M_p}{m_+^3\,R_p^2} 
= 0.9 \times \left(\frac{10\,\mbox{GeV}}{m_+}\right)^{3}\,\left(\frac{M_p}{M_{\rm J}}\right) \,\left(\frac{R_{\rm J}}{R_p}\right)^{2} ~,
\label{eq:opacity2}
\eeqa
where $M_{\rm J} =  1.9 \times 10^{27}~\text{kg}$ and $R_{\rm J} = 7.15\times 10^7$~m represent Jupiter's mass and radius. Note that the charged particle mass is subject to constraints from collider searches, which require $m_+ \gtrsim 500$~GeV~\cite{CMS:2013czn} in the normal vacuum. On the other hand, if charged particles inside the DEP are not in a normal vacuum state, their masses could be different. For instance, if the charged particles in the normal vacuum are some dark baryon state from a dark QCD-like confining gauge interaction, their charged constituent (dark quark) masses could be smaller in the deconfined vacuum state of the DEP, similar in spirit to Refs.~\cite{Bai:2020jfm,Gross:2021qgx}. 

The fraction of star light blocked by a DEP fully occulting its host star is estimated by accounting for the ratio of the cross-sectional areas of the DEP and host star 
\beqa
\label{eq:model_Iblocked}
\frac{\Phi_{\rm blocked}}{\Phi_0} &\sim&  \frac{R_p^2}{R_s^2} \, \tau = \frac{2 \,\alpha^2\,M_p}{m_+^3\,R_s^2}  
= 0.009\times  \left(\frac{10\,\mbox{GeV}}{m_+}\right)^{3}\,\left(\frac{M_p}{M_{\rm J}}\right) \,\left(\frac{R_{\odot}}{R_s}\right)^{2}  ~. 
\eeqa
Here, we used the optical depth for the generalized model in \eqref{eq:opacity2}. We have checked that some other operators coupling DEP particles to the photon---such as the scalar charge radius operator, scalar Rayleigh operator, and fermion magnetic dipole operator---give $\Phi_\text{blocked}/\Phi_0$ too small to be detectable according to the results of Sec.~\ref{sec:mock}.

Another effect that could impact the opacity is the capture of SM matter by the DEP. Captured SM matter may be expected to quickly thermalize with the DEP and sink to the planet's core, in analogy with stellar capture of DM. This may lead to a core that is more opaque than it would be if the DEP is only made of DM, as has been assumed in the calculations above. Thus, the planet may have an opaque core surrounded by a semitransparent ``mantle.'' We neglect this possibility in this work, but it may be worthy of future study. The captured SM matter could also lead to optical emission signals, as studied in \cite{Curtin:2019lhm,Curtin:2019ngc,Kamenetskaia:2022lbf} for the case of dark stars.

\section{Discussion}
\label{sec:discussion}

In this work, we proposed the possibility of a macroscopic dark matter state being an exoplanet of a visible star, which we named ``dark exoplanet'' (DEP). Several possible approaches for the discovery and identification of DEPs are discussed, among which we focus on the transit method for detailed analysis.
Due to its tiny but non-vanishing interaction strength with the SM particles, the DEP may not be completely opaque, rendering a light curve shape distinguishable from that of an ordinary exoplanet (OEP). 
We present the expressions to calculate the transit light curve for a non-opaque exoplanet, and fit the expressions to the measured light curves of two confirmed exoplanets, CoRoT-1~b and K2-44~b, to examine if they can be DEPs.
Though the DEP model does not significantly improve the quality of the fit, it cannot be ruled out, especially for the smaller-radius K2-44~b. Further examinations of mock data show that the DEP transit light curve may not be well-mimicked by an ordinary exoplanet when the DEP is relatively transparent and has a large radius. On the other hand, smaller-radius or larger-opacity DEPs could be hiding in existing transit data masquerading as OEPs. Dark matter models possibly containing such semitransparent macroscopic objects are then discussed.

There are several possible improvements to our work. For simplicity we considered only circular orbits for both the dark and ordinary exoplanets. Though this is not an unreasonable choice, it is common to also fit for the orbital eccentricity $e$ as well as the periastron's orientation $\omega_p$ (see Fig.~\ref{fig:geometry}) from the measured transit light curve. For an eccentric orbit with $\omega_p \neq \pi/2~\text{or}~3\pi/2$, the ingress and egress phases of the transit are asymmetric. The differences between the OEP and DEP light curves are emphasized in one of the stages while diminished in the other. For the cases $\omega_p = \pi/2~\text{or}~3\pi/2$, the ingress/egress shapes are symmetric but will still be different from a circular orbit. The overall effect may not change the conclusion discussed in this work, but it would require additional study by including more model parameters to determine the changes.

There are many other effects we have ignored in our analysis. We do not account for the secondary eclipse of the planet passing behind the star, which may reveal more about the exoplanet properties and help distinguish DEPs from OEPs. Other exoplanet properties we neglect include the presence of a planetary ring, moons, oblateness, atmospheric or topographic features, and night-side emission, among others. We also neglect higher order stellar photometric effects like star spots or stellar rotation and gravity darkening. It is possible that these effects may be confused for opacity effects in a full data analysis. These effects, though small, could become especially important after a DEP candidate is discovered from a simple analysis like the one in this work.

To see a transiting DEP, it must become gravitationally bound to an ordinary star. If stars tend to be born in regions of enhanced dark matter density 
(in analogy to early stars \cite{Abel:2001pr}, though it is unclear if this is true for later-forming stars), then the DEP may become bound {\it in situ}.
Alternatively, the DEP may be captured in a similar way to free-floating planets. While it has been argued (including in the context of Planet Nine \cite{Batygin_2016,P9hyptothesis}, which could itself very speculatively be a DEP) that capture of free-floating planets is rare \cite{Gould,Peter:2009mm,2016ApJ...823L...3L,2016MNRAS.460L.109M,2017MNRAS.472L..75P,Lehmann:2020yxb,Lehmann:2022vdt}, it is no less likely for a free-floating DEP to be captured than a free-floating OEP of a similar mass \cite{Scholtz:2019csj}. Another possibility is if a free-floating DEP passes through a star, its interactions with the star could slow it enough to allow it to become bound. Further study on early DEP--stellar-system formation and DEP capture would help in elucidating the possibility of detecting DEPs and would be necessary for bounds to be set on DEP abundance if they are not detected. 

\vspace{1cm}
\subsubsection*{Acknowledgments}
The authors thank Daniel del Ser and Leo W.H.~Fung for useful discussions. This research has made use of the NASA Exoplanet Archive, which is operated by the California Institute of Technology, under contract with the National Aeronautics and Space Administration under the Exoplanet Exploration Program. The work of YB is supported by the U.S. Department of Energy under the contract DE-SC-0017647. The work of SL is supported by the Area of Excellence (AoE) under the Grant No. AoE/P-404/18-3 issued by the Research Grants Council of Hong Kong S.A.R. The work of NO is supported by the Arthur B. McDonald Canadian Astroparticle Physics Research Institute.  

\appendix

\section{More details on the mock data analysis}
\label{sec:appendix}

\begin{figure}[t!]
\centering
    \includegraphics[width=0.99\textwidth]{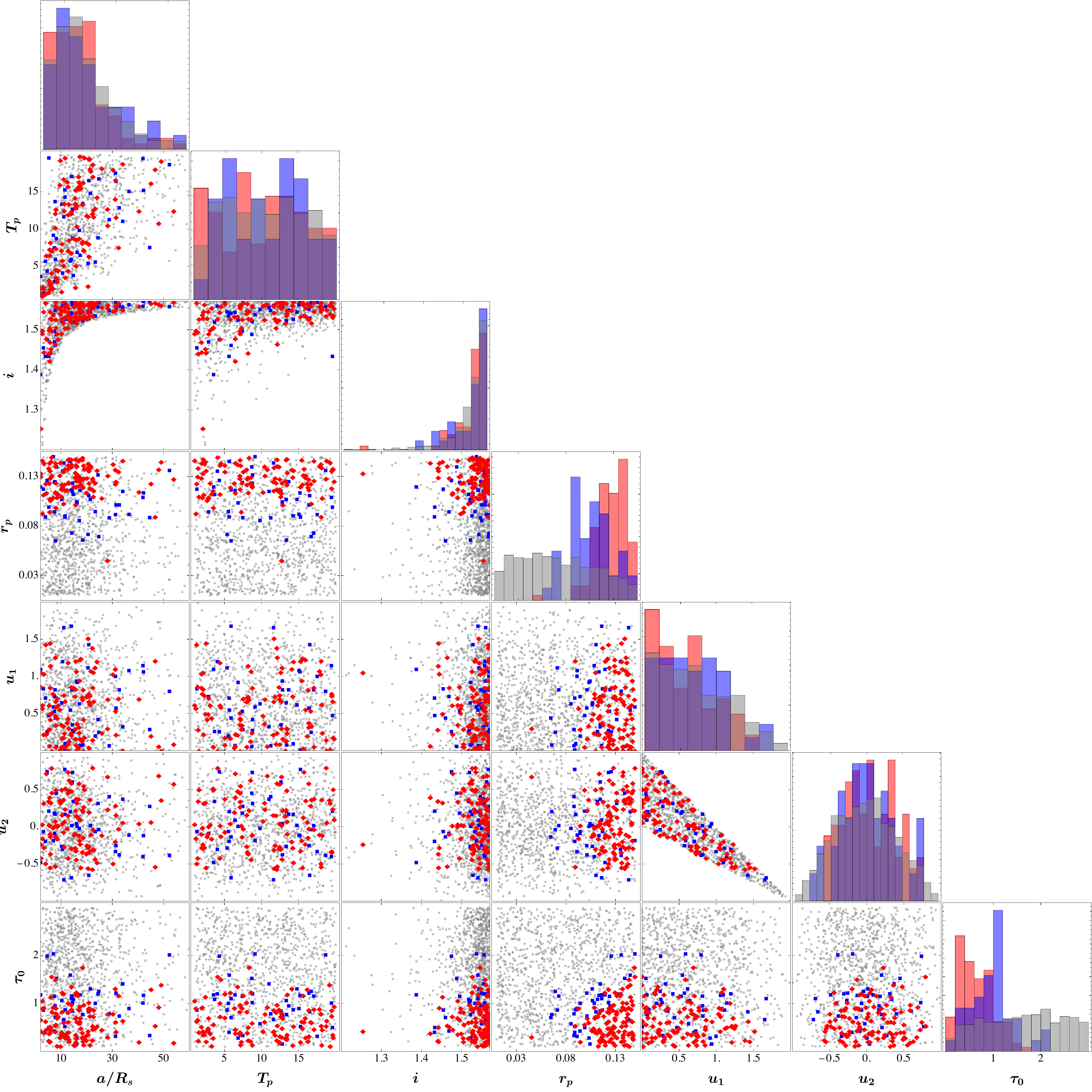}
    \caption{The full corner plot for the DEP and OEP candidates based on DEP mock data, grouped with parameters among $(a, T_p, i, r_p, u_1, u_2, \rnsig)$. 
    The point styles and selection criteria are the same as those in Fig.~\ref{fig:scatter_rp_rnsig}, except with the green dots remaining gray here.
    The left panel of Fig.~\ref{fig:scatter_rp_rnsig} can be found in the bottom row, fourth column, and the top-right panel of Fig.~\ref{fig:scatter_rp_rnsig} is in the second row, first column.
    }
    \label{fig:mock_cornerplot}
\end{figure}

In this appendix, we provide additional analysis of the mock data presented in Fig.~\ref{fig:scatter_rp_rnsig}. A full corner plot of the mock DEP data parameters, with those points that can statistically distinguish between DEPs and OEPs, is shown in Fig.~\ref{fig:mock_cornerplot}. Note that the left and top-right panels of Fig.~\ref{fig:scatter_rp_rnsig} are contained within this plot (except the green circles of Fig.~\ref{fig:scatter_rp_rnsig} are simply shown as gray circles here). 

\begin{figure}[t!]
\centering
    \includegraphics[width=0.5\textwidth]{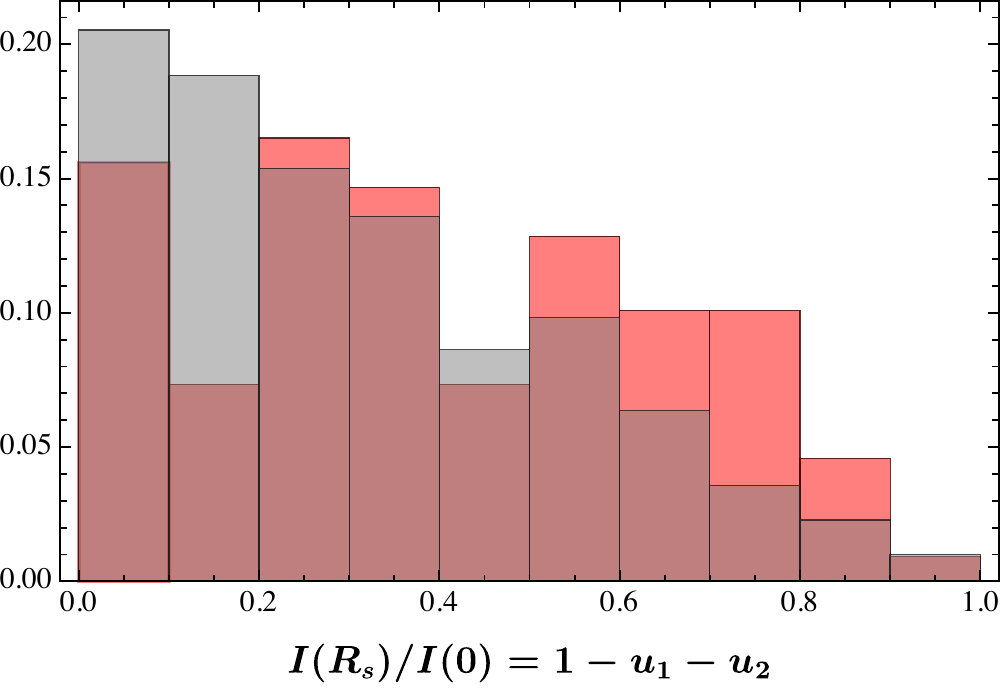}
    \caption{The probability distribution of $I(R_s)/I(0)=1-u_1-u_2$ for the DEP $\alpha = 0.01$ (red) and OEP (gray) candidate mock data. The DEP $0.01 < \alpha < 0.05$ candidates (blue points in previous figures) are not plotted for easier readability.
    }
    \label{fig:mock_histo_IRs}
\end{figure}

In addition to the DEP candidate preferences on $r_p$, $\rnsig$, and $i$ noted in the main text, there is also in Fig.~\ref{fig:mock_cornerplot} a preference compared to the OEP candidates for the DEP candidates to have $u_2$ not too close to its maximum or minimum $\pm 1$. For such values, $I(R_s)/I(0) = 1 - u_1 - u_2$ is close to zero, so it may be easier to confuse limb darkening and DEP opacity effects for such stars. Indeed, the histogram in Fig.~\ref{fig:mock_histo_IRs} for the $I(R_s)/I(0)$ distribution shows a preference for larger values of this parameter in the DEP candidates compared to the OEP candidates, indicating a preference for stars that are more uniformly bright.

\begin{figure}[t!]
\centering
    \includegraphics[width=0.5\textwidth]{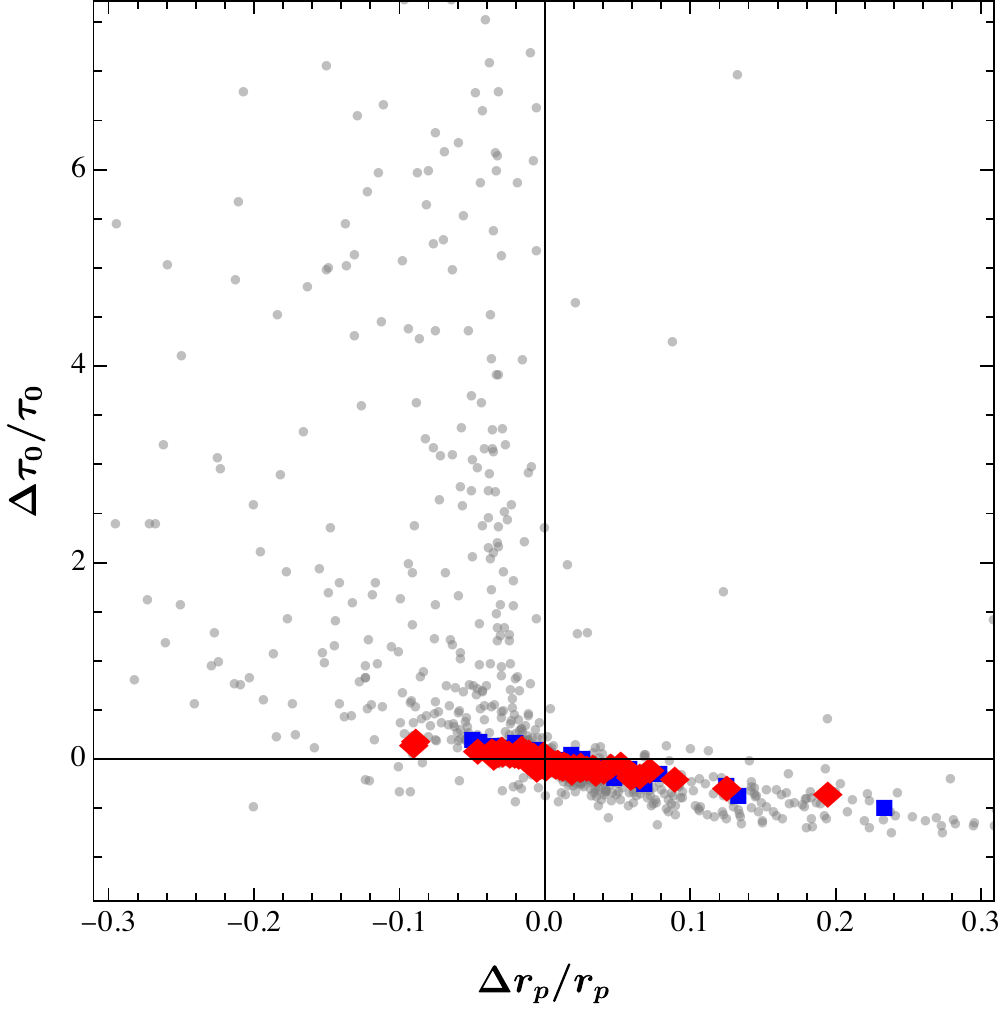}
    \caption{Plot for the mock data presented in Fig.~\ref{fig:mock_cornerplot} showing the fractional difference between the generated and fitted DEP parameters. Positive (negative) values indicate the fitted value is greater (smaller) than the generated ``true'' value.}
    \label{fig:mock_rp_rnsig_percentdiff}
\end{figure}

Fig.~\ref{fig:mock_rp_rnsig_percentdiff} shows a plot of the fractional difference between the fitted and generated DEP parameters for $r_p$ and $\rnsig$.  Note that the parameters $r_p$ and $\rnsig$ are inversely correlated in their errors. Intuitively, when $\rnsig$ is overestimated, $r_p$ tends to be underestimated and vice versa because the fractional dimming of the star must be held constant. Indeed, as explained in the main text, when $\rnsig < 1$, the best-fitting regions obey $\tau_0 \propto r_p^{-2}$. For the gray points, when $\rnsig \gg 1$ the points are essentially degenerate in $\rnsig$, so some points can have large fractional differences for their best-fit points (some of these are cut off by the plot range).

\providecommand{\href}[2]{#2}\begingroup\raggedright\endgroup

\end{document}